\documentclass{mn2e}

\input graphicx.tex

\newcommand{\OIII}{[O~{\sc iii}]}
\newcommand{\OII}{[O~{\sc ii}]}
\newcommand{\NIII}{[N~{\sc iii}]}
\newcommand{\NII}{[N~{\sc ii}]}
\newcommand{\SII}{[S~{\sc ii}]}
\newcommand{\SIII}{[S~{\sc iii}]}
\newcommand{\SIV}{[S~{\sc iv}]}
\newcommand{\HII}{H~{\sc ii}}

\newcommand{\eq}{\begin{equation}}
\newcommand{\en}{\end{equation}}

\def\ltsima{$\; \buildrel < \over \sim \;$}
\def\simlt{\lower.5ex\hbox{\ltsima}}
\def\gtsima{$\; \buildrel > \over \sim \;$}
\def\simgt{\lower.5ex\hbox{\gtsima}}

\title[\OIII/\NII\ as an abundance indicator]
      {\OIII/\NII\ as an abundance indicator at high redshift}  
      
\author[M. Pettini and B.E.J. Pagel]
       {Max  Pettini$^{1}$ and Bernard E. J. Pagel$^{2}$\\
       $^{1}$ Institute of Astronomy, Madingley Road, Cambridge CB3 0HA \\ 
       $^{2}$ Astronomy Centre, School of Science \& Technology, Sussex University, Brighton BN1 9QH \\}

\date{Received ........; in original form 2003 Dec.............}

\begin{document}

\maketitle

\begin{abstract}
Among `empirical' methods of estimating oxygen abundances in 
extragalactic \HII\ regions, the use of the ratio of nebular 
lines of \OIII\ and \NII, first introduced by Alloin et al. 
in 1979, is reappraised with modern calibration data and 
shown to have certain advantages over 
$R_{23} \equiv$ (\OII\ + \OIII)/H$\beta$ 
and $N2 \equiv$ \NII~$\lambda 6583$/H$\alpha$,
particularly when applied to star-forming galaxies at
high redshifts.
\end{abstract}


\section{Introduction}
Extragalactic \HII\ regions are central to the study of chemical
abundances in the universe (e.g. Garnett 2004).
Accurate abundance measurements require the determination
of the electron temperature, which in turn is usually obtained from
the ratios of  auroral to nebular line intensities, 
such as  \OIII~$\lambda 4363/\lambda 5007$.
This is often referred to as the `direct' $T_{\rm e}$ method.
A well-known difficulty, however, arises from the fact that,
as the metallicity increases, the electron 
temperature decreases (since the cooling is via metal lines)
and the auroral lines eventually become too faint to measure. 
Detailed modelling can overcome this
problem. For example, Kewley \& Dopita (2002) 
have recently used a combination of stellar population 
synthesis and photoionization models to develop a set of 
ionization parameter and abundance
diagnostics based only on the 
strong optical emission lines.
Their `optimal' method uses ratios 
of \NII, \OII, \OIII, \SII, \SIII\ and the Balmer lines,
that is the full complement of strong nebular
lines accessible from the ground. 
They also recommended procedures 
for the derivation of abundances in cases
where only a subset of these lines is available.

The advent of 8-10\,m class telescopes has made it
possible to extend observations of these lines to galaxies
at high redshifts (e.g. Teplitz et al. 2000;
Pettini et al. 2001; Kobulnicky et al. 2003; 
Lilly, Carollo, \& Stockton 2003;
Steidel et al. 2004).
At high redshift, however, the study of nebular
emission encounters new obstacles.
The first and obvious difficulty is that,
given the much larger distances involved,
the line fluxes are reduced dramatically.
The problem is compounded by the fact that
the lines are redshifted into the near-infrared
wavelength region, where the sky background is
orders of magnitude higher than in the optical.
Second, at a given redshift only a subset of the
strong lines will fall within an atmospheric window 
and thus be accessible from the ground. 
Even at particularly favourable
redshifts, such as $z \simeq 2.3$ which shifts
\OII, \OIII\ and H$\beta$, and H$\alpha$ and \NII\
respectively into the middle range of the 
$J$, $H$ and $K$ bands, it is not possible to
record all these lines in a single exposure.
Rather, different spectrograph settings are normally
required, and this can easily introduce additional 
errors in the relative flux calibrations.
For all of these reasons, there are practical 
limitations to the accuracy with which emission
line ratios can be measured in high redshift objects.


\begin{figure*}
        \centering
        \includegraphics[height=0.85\textwidth,angle=270]{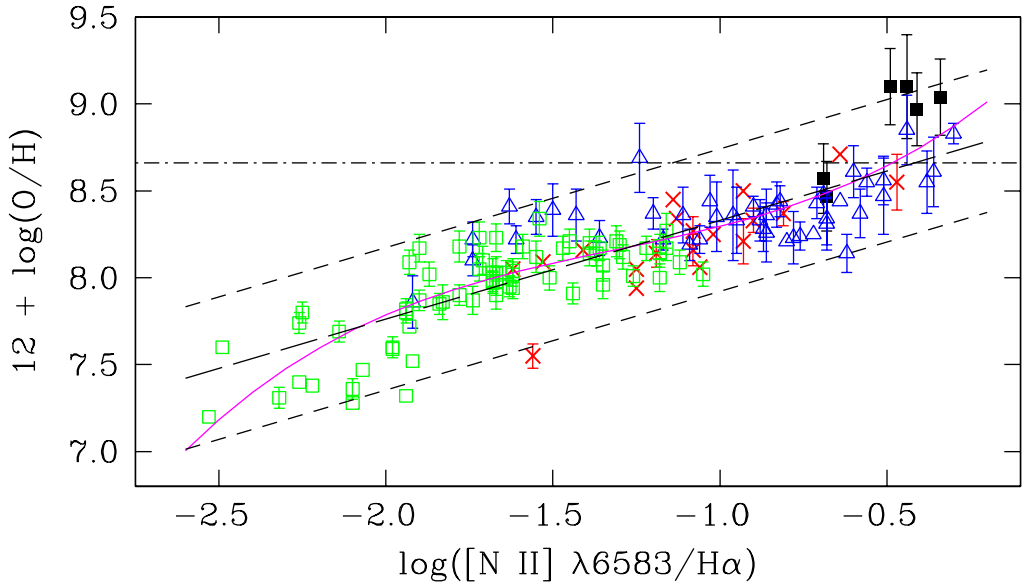}
\caption{Oxygen abundance against the $N2$ index in extragalactic \HII\ regions.
The symbols have the 
following meanings: {\it Filled squares} -- (O/H) from photoionisation
models by D\'{\i}az et al. (1991 -- M~51) and 
Castellanos, D\'{\i}az, \& Terlevich (2002a,b -- NGC~925 and NGC~1637).
{\it Crosses} -- M~101 (Kennicutt et al. 2003).
{\it Open triangles} -- spiral and irregular galaxies
(Castellanos et al. 2002a; Garnett et al. 1997; Gonz{\' a}lez-Delgado et al. 1994, 1995;
Kobulnicky \& Skillman 1996;  Mathis, Chu, \& Peterson 1985; 
Pagel, Edmunds, \& Smith 1980;  Pastoriza et al. 1993; 
Shaver et al. 1983 [30 Dor and NGC 346]; Skillman, C{\^ o}t{\' e}, \& Miller 2003;
V\'{\i}lchez et al. 1988).  
{\it Open squares} -- blue compact galaxies (Campbell, Terlevich, \& Melnick 1986; 
Garnett 1990; Izotov, Thuan, \&  Lipovetsky 1994; 
Kniazev et al. 2000; Kunth \& Joubert 1985;  
Pagel  et al. 1992; Skillman \& Kennicutt 1993; Skillman et al. 1994;
Thuan, Izotov, \&  Lipovetsky 1995; van Zee 2000; Walsh \& Roy 1989).
Error bars in $\log {\rm (O/H)}$ are only plotted when they are
larger than the symbol to which they refer.
The long-dash line is the best fitting linear relationship:
$12 + \log {\rm (O/H)} = 8.90 + 0.57 \times N2$. 
The short-dash lines encompass 95\% of the measurements
and correspond to a range in $\log {\rm (O/H)} = \pm 0.41$
relative to this linear fit.
The continuous line is a cubic function of the form
$12 + \log {\rm (O/H)} = 9.37 + 2.03 \times N2 + 1.26 \times N2^{2} + 0.32 \times N2^{3}$
which, however, gives only a slightly better fit to the data
(95\% of the data points are within $\pm 0.38$ of this line).
The dot-dash horizontal line shows the solar oxygen abundance
$12 + \log {\rm (O/H)} = 8.66$ 
(Allende-Prieto, Lambert, \& Asplund 2001; Asplund et al. 2004).
}
\end{figure*}

There is thus an incentive to explore simpler
abundance indicators based on only a few emission
lines---preferably close in wavelength---which, 
while admittedly less accurate than the full 
treatments mentioned above, may still be 
adequate to characterise the chemical enrichment
of distant galaxies. The idea of such indicators goes 
back to a fundamental paper by Searle (1971) who, 
noting some measurements of nebular \OIII\ lines 
in M33 by Aller (1942), himself found a systematic increase 
in \OIII/H$\beta$ and a corresponding decrease in 
\NII/H$\alpha$ with galactocentric distance as a generic 
property of giant \HII\ regions in late-type spiral galaxies.
Searle pointed out that these trends were most likely due to 
radial abundance gradients, thereby
raising the possibility of using these line ratios
more generally as indicators of abundance.    
Such indicators can be calibrated against
more secure abundance determinations, based on 
auroral line measurements or on 
detailed photoionization models, 
to establish their typical accuracy;
for this reason they are often referred to as
`empirical' methods. 
The first exploration of such a method 
was by Jensen, Strom \& Strom (1976)
who considered the ratio \OIII/H$\beta$.
The subsequent extension of this idea 
by Pagel et al. (1979), who included \OII,
led to the most widely used
abundance indicator, the $R_{23}$ index
which relates the abundance of oxygen to the 
ratio of (\OII\ + \OIII) to H$\beta$.
In the last 25 years there have been
many efforts to refine the calibration of 
$R_{23}$.
The most successful are 
the calibration by McGaugh (1991) which is based on 
photoionisation models and 
the empirical one by Pilyugin (2003); both 
improve the accuracy by making use of the 
\OIII/\OII\ ratio as an ionisation parameter.
Even so, systematic differences of up to 0.5\,dex
remain between the oxygen abundances derived from
the $R_{23}$ index and those deduced from
the `direct' $T_{\rm e}$ method 
(e.g. Skillman, C{\^ o}t{\' e}, \& Miller 2003;
Kennicutt, Bresolin, \& Garnett 2003).
When compounded with the well-known double-valued
behaviour of (O/H) vs. $R_{23}$, this can lead to 
order of magnitude uncertainties in (O/H) 
in galaxies at $z \simeq 3$ (Pettini et al. 2001).

Partly to circumvent this latter problem, analogous
$S_{23} \equiv$ (\SII\ + \SIII)/H$\beta$ 
(V\'{\i}lchez \& Esteban 1996;
Christensen, Petersen, \& Gammelgaard 1997;
D\'{\i}az \& P\'{e}rez-Montero 2000)
and $S_{234} \equiv$ (\SII\ + \SIII + \SIV)/H$\beta$
(Oey et al. 2002) indices have been proposed.
Although these sulphur-line based indices turn
over at higher metallicities than $R_{23}$,
the sulphur lines are weaker than
their oxygen counterparts and their
red and infrared rest wavelengths place them beyond
the reach of ground-based observations at redshifts
$z \simgt 1.5$.

More promising are methods which involve \NII\ and H$\alpha$.
In this paper we reconsider the $N2$ index
($N2 \equiv$ log\{\NII~$\lambda 6583$/H$\alpha$\})
recently discussed by Denicol{\' o}, Terlevich, \& Terlevich 
(2002), and show how its accuracy can be further improved in 
the high abundance regime by comparison with \OIII~$\lambda 5007$/H$\beta$.

\section{The $N2$ index}


\begin{figure*}
        \centering
        \includegraphics[height=0.85\textwidth,angle=270]{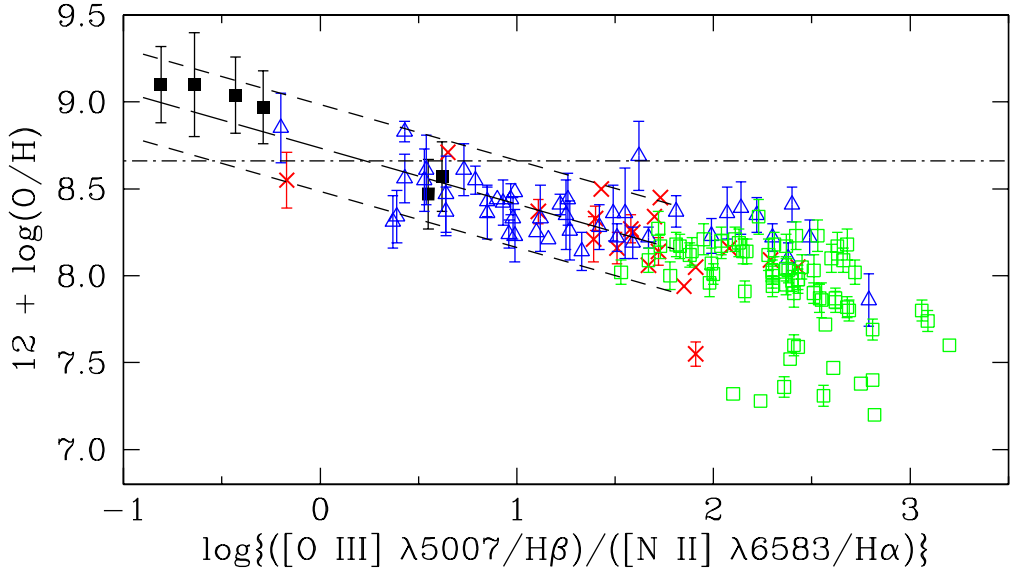}
\caption{Oxygen abundance against the $O3N2$ index in extragalactic \HII\ regions.
The symbols have the same meaning as in Figure 1.
The long-dash line is the best fitting linear relationship,
$12 + \log {\rm (O/H)} = 8.73 - 0.32 \times O3N2$, 
valid when $O3N2 < 1.9$
The short-dash lines encompass 95\% of the measurements which
satisfy this condition
and correspond to a range in $\log {\rm (O/H)} = \pm 0.25$.
The dot-dash horizontal line shows the solar oxygen abundance
$12 + \log {\rm (O/H)} = 8.66$ 
(Allende-Prieto, Lambert \& Asplund 2001; Asplund et al. 2004).
}
\end{figure*}

Following up on earlier work by 
Storchi-Bergmann, Calzetti, \& Kinney (1994)
and by Raimann et al. (2000), 
Denicol{\' o} et al. (2002) have focussed attention 
again on the $N2$ index, pointing out its
usefulness in the search for low metallicity galaxies.
The \NII/H$\alpha$ ratio is highly sensitive to the
`metallicity', as measured by the oxygen abundance (O/H),
through a combination of two effects. 
As (O/H) decreases below solar, 
there is a tendency for the ionization to 
increase (either from hardness of the ionizing spectrum 
or from the ionization parameter, or both), decreasing the 
ratio \NII/\NIII,
and, furthermore, the (N/O) ratio itself decreases 
at the high-abundance end  
because of the secondary nature of nitrogen
(e.g. Henry, Edmunds, \& K{\" o}ppen 2000;
Pettini et al. 2002a).
From the ground, H$\alpha$ and \NII\ can be 
followed all the way to redshift
$z \simeq 2.5$ (the limit of the $K$ band).
With thousands of galaxies now known 
at $z > 1$ (Madgwick et al. 2003; 
Steidel et al. 2004; Abraham et al., in preparation),
the $N2$ index offers the means to determine
metallicities in a wholesale manner over 
look-back times which span most
of the age of the universe.  
It is thus worthwhile reconsidering its accuracy.

Denicol{\' o} et al. (2002) compiled 
an extensive sample of nearby extragalactic 
\HII\ regions which could be used to calibrate 
$N2$ vs. (O/H).   
We have revised their database by:
(i) including only \HII\ regions with values
of (O/H) determined either via the $T_{\rm e}$ method
or with detailed photoionization modelling, and 
 excluding the untabulated estimates
for \HII\ regions from the catalogue of 
Terlevich et al. (1991), because for many of these
(O/H) had been estimated from `empirical' 
(i.e. $R_{23}$ or $S_{23}$) indices;
(ii) updating it with recent measurements by Kennicutt et al. 
(2003) for \HII\ regions in M~101 and by Skillman et al. 
(2003) for dwarf irregular galaxies in the Sculptor Group.

The total sample, shown in Figure 1, 
consists of 137 extragalactic \HII\
regions with well determined values of (O/H) and 
$N2$. In all but six cases (shown with filled
symbols) the oxygen abundance was determined with
the $T_{\rm e}$ method. References to the original
works are given in the Figure caption. 
It can be readily appreciated from Figure 1
that, while $N2$ does indeed increase monotonically
with $\log {\rm (O/H)}$, a linear relationship is only
an approximate representation of the data, which rather 
tend to lie along an S-shaped curve.
In particular, when (O/H)\,$\simeq 8.2 \pm 0.2$,
the value of $N2$ can span one order of magnitude,
from $N2 \simeq -1.8$ to $N2 \simeq -0.8$,
whereas at solar metallicity and beyond 
\NII\ tends to saturate (Baldwin, Phillips \& Terlevich 1981)
because it comes to dominate the cooling 
(Kewley \& Dopita 2002).
Forcing a linear fit to the sample,\footnote{We performed an 
unweighted least squares fit to the data because in compilations 
such as this the error estimates are highly heterogeneous. However,
a weighted least squares fit gives very similar results.}
we obtain a line of best fit according to the relation:
\begin{equation}
12 + \log {\rm (O/H)} = 8.90 + 0.57 \times N2
\end{equation}
shown with the long-dash line in Figure 1.
Both the slope and the intercept are lower than
the values of the fit by 
Denicol{\' o} et al. (2002) who proposed
$12 + \log {\rm (O/H)} = 9.12 + 0.73 \times N2$.
While the formal statistical errors on the slope 
and intercept are small (0.03 and 0.04 respectively),
of more interest is the dispersion of the points
about the line of best fit in Figure 1. Specifically,
95\% (68\%) of the measurements of $\log {\rm (O/H)}$ lie 
within $\pm 0.41$ ($\pm 0.18$) of the line defined by eq. (1); 
the $2 \sigma$ limits are shown as short-dash lines in Figure 1.
Only a marginally better fit to the data is provided by
a third order polynomial of the form:
\begin{equation}
12 + \log {\rm (O/H)} = 9.37 + 2.03 \times N2 + 1.26 \times N2^{2} + 0.32 \times N2^{3}
\end{equation}
(valid in the range $-2.5 < N2 < -0.3$),
with 95\% (68\%) of the measurements being within 
$\pm 0.38$ ($\pm 0.18$) of the values given by eq. (2).
This cubic fit is indicated by the continuous line in Figure 1.
We conclude that with the $N2$ calibrator
it is possible to estimate
the abundance of oxygen to within a factor of $\sim 2.5$
at the 95\% confidence level. This accuracy is comparable
to that of the $R_{23}$ method.

\section{The $O3N2$ Index}

We now consider to what degree the accuracy in the
determination of (O/H) can be improved by considering
{\it two} ratios, \NII/H$\alpha$ and \OIII/H$\beta$. 
Alloin et al. (1979) were the first to 
introduce the quantity 
$O3N2 \equiv \log $\{([O~{\sc iii}]~$\lambda 5007$/H$\beta$)/
([N~{\sc ii}]~$\lambda 6583$/H$\alpha$)\},\footnote{This
definition is slightly different from the original one proposed by
Alloin et al. (1979) who included both \OIII\ doublet lines in the
numerator of the first ratio.}
but since then the $O3N2$ index has been 
comparatively neglected in nebular abundance studies.
There is now a sufficient body of high quality data to reassess its merits.
We expect the inclusion of \OIII\ to be most useful in the
high metallicity regime where \NII\ saturates 
but the strength of \OIII\ continues
to decrease with increasing metallicity.

Figure 2 shows how $O3N2$ varies with (O/H) for the 137
extragalactic \HII\ regions in our sample.
Clearly this method is of little use when 
$O3N2 \simgt 2$, but at lower values there appears to be a
relatively tight, linear and steep relationship
between $O3N2$ and $\log {\rm (O/H)}$. 
A least squares linear fit to
the data in the range $-1 < O3N2 < 1.9$ yields the relation:
\begin{equation}
12 + \log {\rm (O/H)} = 8.73 - 0.32 \times O3N2
\end{equation}
which is shown by the long-dash line in Figure 2,
with 95\% (68\%) of the measurements within 0.25 (0.14) dex of the
best fit line; the short-dash lines in Figure 2 show the
$2 \sigma$ limits.
The caveats are that the statistics are somewhat limited,
with only 65 out of the 137 \HII\ regions in our sample
satisfying the condition $O3N2 < 1.9$, and that the 
relationship defined by eq. (3) relies heavily on the 
four data points with $12 + \log {\rm (O/H)} \simeq 9.0$
deduced from photoionisation models rather than the 
`direct' $T_{\rm e}$ method.
Nevertheless, in the high metallicity regime the 
$O3N2$ method appears promising and it would
be very worthwhile extending the database of nearby 
\HII\ region measurements to improve its statistics.

\section{Conclusions}

In this note we have reassessed the usefulness of 
`empirical' methods based on the \NII/H$\alpha$ ratio
for determining the oxygen abundance in \HII\ regions,
particularly with an eye to their application
to the analysis of star-forming galaxies at high redshift.
We have revised the extensive compilation prepared
by Denicol\' o et al. (2002) to include only \HII\ regions
where the oxygen abundance is believed to be known reliably,
either because the electron temperature has been determined 
with high precision,
or through detailed modelling of the ionisation of the nebula.
Our main conclusions are as follows.

1.  The relationship between 
$N2 \equiv \log$ (\NII~$\lambda 6583$/H$\alpha$) 
and $\log {\rm (O/H)}$ is only approximately linear and yields 
(with the revised data set considered here)   
estimates of (O/H) which are only accurate to within 
$\sim 0.4$\,dex at the 95\% confidence level.
This level of precision is comparable to that of
the more widely used $R_{23}$ method.

2. At moderate to high metallicities, greater 
than $\sim 1/4$ solar, the oxygen abundance
can be deduced to within a factor of $\sim 0.25$\,dex
(again at the 95\% confidence level) if
$O3N2 \equiv \log $\{([O~{\sc iii}]~$\lambda 5007$/H$\beta$)/([N~{\sc ii}]~$\lambda 6583$/H$\alpha$)\}
is lower than $\sim 1.9$. 
The $O3N2$ index is particularly useful at solar and 
super-solar metallicities, where $N2$ saturates.

When applied to high redshift galaxies, the 
$N2$ and $O3N2$ indices offer several advantages
compared to the more familiar $R_{23}$ method.
First, the N-based indices have a monotonic
behaviour with (O/H), overcoming the ambiguities
introduced by the double-valued character of $R_{23}$.
Second, both $N2$ and $O3N2$ rely on ratios of
emission lines which are close in wavelength,
so that corrections for reddening, and accurate
flux calibration of the spectra, are not necessary.
Third, both indices are highly sensitive to
the oxygen abundance and are thus particularly
well suited to the analysis of data of only 
moderate signal-to-noise ratio. 
Specifically, the slope of 0.57 of the relationship
between $\log {\rm (O/H)}$ and $N2$ implies that one only needs
to measure the \NII/H$\alpha$ ratio with an accuracy
of about 25\%. Better precision would not improve
the estimate of (O/H) since the random error would
then be of minor importance relative to the systematic
uncertainty of $\pm 0.4$\,dex in the method.
Similar considerations apply to the $O3N2$ index which
is inherently more accurate in the high metallicity
regime and varies even more steeply with $\log {\rm (O/H)}$.

The high sensitivity of $O3N2$ to the oxygen abundance
also helps to mitigate another complication which may
arise preferentially at high redshift. 
The galaxies accessible to detailed spectroscopic studies
at redshifts $z = 1 - 3$ support very high rates of star
formation, one to two orders of magnitude higher than those
seen in the local universe (e.g. Steidel et al. 2003, 2004).
There are also mounting indications that, even at these early
epochs, such galaxies have already reached 
near-solar metallicities (e.g. Pettini 2004).
When galactic chemical evolution proceeds at such a fast
pace, the production of primary nitrogen from intermediate
and low mass stars can lag behind that of oxygen from Type II
supernovae, as found by Pettini et al. (2002b) in the 
$z = 2.7276$ Lyman break galaxy MS1512-cB58 
(see also Matteucci \& Pipino 2002).
How far (N/O) is `out of equilibrium' depends
on the metallicity, the star formation rate, and the yet
poorly constrained timescale for the release of primary nitrogen
(e.g. Meynet \& Pettini 2003).
However, even in cases where
N is over-deficient relative to O by factors of 2--3, 
as in MS1512-cB58, 
the slope of $-0.32$ in the relation 
$\log {\rm (O/H)}$ vs.  $O3N2$ limits the error in 
$\log {\rm (O/H)}$ to $\simlt 0.15$ 
(given that (\NII/H$\alpha$) varies 
linearly or slower with the nitrogen abundance 
in this regime).

Finally, it is encouraging 
that abundance calibrators developed from measurements
in individual extragalactic \HII\ regions seem to be
fairly robust when applied to the integrated light
of galaxies (Kobulnicky, Kennicutt, \& Pizagno 1999;
Moustakas \& Kennicutt, in preparation).
Thus, we propose that, with the aid of the $O3N2$ index,
it should be possible to measure the oxygen abundances
of star forming galaxies at $z > 1$
to better than a factor of two with only a relatively modest
investment in observing time.\\

We are grateful to Chris Akerman, Mike Irwin and Samantha Rix
for help with the production of the figures, and to Mike Dopita,
Lisa Kewley and Chuck Steidel for valuable comments.

\bsp
\end{document}